\documentclass[prd,twocolumn,showpacs,preprintnumbers,amsmath,nofootinbib,amssymb,floatfix]{revtex4-1}
\usepackage{graphicx}
\usepackage{epsfig,float}
\usepackage[sort&compress]{natbib}
\usepackage{subfigure}
\usepackage{amsmath}
\usepackage{amsfonts}
\usepackage{cancel}
\usepackage{lmodern,dsfont}
\usepackage{amssymb}
\usepackage{hyperref}
\begin{document}
\arraycolsep1.5pt
\newcommand{\Ima}{\textrm{Im}}
\newcommand{\Rea}{\textrm{Re}}
\newcommand{\mev}{\textrm{ MeV}}
\newcommand{\gev}{\textrm{ GeV}}
\newcommand{\dtres}{d^{\hspace{0.1mm} 3}\hspace{-0.5mm}}
\newcommand{\rts}{ \sqrt s}
\newcommand{\non}{\nonumber \\[2mm]}
\newcommand{\eps}{\epsilon}
\newcommand{\half}{\frac{1}{2}}
\newcommand{\thalf}{\textstyle \frac{1}{2}}
\newcommand{\Nmass}{M_{N}} 
\newcommand{\delmass}{M_{\Delta}} 
\newcommand{\pimass}{\mu}  
\newcommand{\rhomass}{m_\rho} 
\newcommand{\piNN}{f}      
\newcommand{\rhocoup}{g_\rho} 
\newcommand{\fpi}{f_\pi} 
\newcommand{\f}{f} 
\newcommand{\nucfld}{\psi_N} 
\newcommand{\delfld}{\psi_\Delta} 
\newcommand{\fpiNN}{f_{\pi N N}} 
\newcommand{\fpiND}{f_{\pi N \Delta}} 
\newcommand{\GMquark}{G^M_{(q)}} 
\newcommand{\vecpi}{\vec \pi}
\newcommand{\vectau}{\vec \tau}
\newcommand{\vecrho}{\vec \rho}
\newcommand{\delmu}{\partial_\mu}
\newcommand{\delMu}{\partial^\mu}
\newcommand{\nn}{\nonumber}
\newcommand{\bi}{\bibitem}
\newcommand{\vs}{\vspace{-0.20cm}}
\newcommand{\be}{\begin{equation}}
\newcommand{\ee}{\end{equation}}
\newcommand{\ba}{\begin{eqnarray}}
\newcommand{\ea}{\end{eqnarray}}
\newcommand{\ropi}{$\rho \rightarrow \pi^{0} \pi^{0}
\gamma$ }
\newcommand{\roeta}{$\rho \rightarrow \pi^{0} \eta
\gamma$ }
\newcommand{\omepi}{$\omega \rightarrow \pi^{0} \pi^{0}
\gamma$ }
\newcommand{\omeeta}{$\omega \rightarrow \pi^{0} \eta
\gamma$ }
\newcommand{\ul}{\underline}
\newcommand{\del}{\partial}
\newcommand{\rth}{\frac{1}{\sqrt{3}}}
\newcommand{\rsix}{\frac{1}{\sqrt{6}}}
\newcommand{\sq}{\sqrt}
\newcommand{\fr}{\frac}
\newcommand{\pr}{^\prime}
\newcommand{\ov}{\overline}
\newcommand{\Gm}{\Gamma}
\newcommand{\rw}{\rightarrow}
\newcommand{\rgl}{\rangle}
\newcommand{\De}{\Delta}
\newcommand{\Dp}{\Delta^+}
\newcommand{\Dm}{\Delta^-}
\newcommand{\Dz}{\Delta^0}
\newcommand{\Dpp}{\Delta^{++}}
\newcommand{\Sg}{\Sigma^*}
\newcommand{\Sp}{\Sigma^{*+}}
\newcommand{\Sm}{\Sigma^{*-}}
\newcommand{\Sz}{\Sigma^{*0}}
\newcommand{\X}{\Xi^*}
\newcommand{\Xm}{\Xi^{*-}}
\newcommand{\Xz}{\Xi^{*0}}
\newcommand{\Om}{\Omega}
\newcommand{\Omm}{\Omega^-}
\newcommand{\kp}{K^+}
\newcommand{\kz}{K^0}
\newcommand{\pip}{\pi^+}
\newcommand{\pim}{\pi^-}
\newcommand{\piz}{\pi^0}
\newcommand{\et}{\eta}
\newcommand{\kb}{\ov K}
\newcommand{\km}{K^-}
\newcommand{\kbz}{\ov K^0}
\newcommand{\ksb}{\ov {K^*}}

\newcommand{\re}{\text{Re}}
\newcommand{\im}{\text{Im}}

\def\tstrut{\vrule height2.5ex depth0pt width0pt} 
\def\jtstrut{\vrule height5ex depth0pt width0pt} 

\title{On the isospin 0 and 1 resonances
from $\pi \Sigma$ photoproduction data}

\author{L. Roca$^1$ and E. Oset$^2$}
\affiliation{
$^1$ Departamento de F\'\i sica, Universidad de Murcia, E-30100 Murcia, Spain.
$^2$ Departamento de F\'{\i}sica Te\'orica and IFIC, Centro Mixto Universidad de Valencia-CSIC,
Institutos de Investigaci\'on de Paterna, Aptdo. 22085, 46071 Valencia,
Spain
 }

\date{\today}

\begin{abstract}

Recently we presented a successful strategy to extract the position of the two 
$\Lambda(1405)$ poles from experimental photoproduction data on the
$\gamma p \to K^+ \pi^0 \Sigma^0 $ reaction at Jefferson Lab.
Following a similar strategy, we extend  the previous method to incorporate also the
isospin 1 component which allows  us to consider in addition
the experimental data on $\gamma p
\to K^+ \pi^\pm \Sigma^\mp$. The idea is based on considering a production mechanism as
model  independent as possible and implementing the final state interaction of the final
meson-baryon pair 
based on small modifications of the unitary chiral perturbation theory
amplitudes. Good fits to the data are obtained with this procedure,
by means of which we can also predict the cross sections for the $K^- p
\to \bar K N$, $\pi \Sigma$, $\pi\Lambda$ reactions for the different charge
channels. Besides the two poles found for the $\Lambda(1405)$ resonance, we discuss the
possible existence of an isospin 1 resonance in the vicinity of the $\bar K N$
threshold.

\end{abstract}


\maketitle

\section{Introduction}
\label{Intro}

In a recent paper \cite{Roca:2013av} we analyzed the CLAS data for photoproduction of the $\Lambda(1045)$ \cite{Moriya:2013eb} in the reaction 
$\gamma p\to K^+ \pi^0\Sigma^0$, which filters I=0 for the $\pi^0\Sigma^0$ state, where clear peaks where seen related to the $\Lambda(1045)$ excitation. The aim was to determine the mass and width of the two $\Lambda(1405)$ states, predicted by all the latest works based on chiral dynamics. While the nature  of the $\Lambda(1045)$ as generated from the interaction of meson baryon channels with strangeness S=-1 has been long accepted \cite{Dalitz:1960du,Dalitz:1967fp,Veit:1984an}, the use of chiral dynamics and unitary schemes brought new light into this issue \cite{Kaiser:1995eg,Kaiser:1996js,Oset:1998it,Oller:2000fj,Lutz:2001yb,Oset:2001cn,Hyodo:2002pk,cola,GarciaRecio:2002td,GarciaRecio:2005hy,Borasoy:2005ie,Oller:2006jw,Borasoy:2006sr,hyodonew,kanchan,hyodorev,ollerguo}, and the $\Lambda(1045)$ appears in all these works by using chiral Lagrangians and adjusting a minimum amount of parameters to reproduce $\bar K N$ data. Hints of the existence of two, rather than one states were found in \cite{Fink:1989uk,Oller:2000fj} and a thorough study of the existence of two poles was conducted in \cite{cola}. Since then, all the new works on chiral dynamics obtain two poles and this has come as a broadly accepted fact, even reflected in the PDG \cite{pdg}.

  All works on chiral dynamics fit $\bar K N$ data to determine the few
free parameters of the theory and then determine the position of the
poles. Some also consider data on $\Lambda(1405)$ production, like
\cite{Oller:2006jw}. The omission of $\Lambda(1405)$ production data in
most works was justified because of the nontrivial dynamics of the
reaction process, although work in this direction has been done
\cite{Kaiser:1996js,Nacher:1999ni,Hyodo:2003jw,magaslam,geng,Borasoy:2007ku}.
However, the novelty of \cite{Roca:2013av} was to show that the
$\Lambda(1405)$ photoproduction data by themselves had the capacity to
provide the pole positions of the two $\Lambda(1405)$ states which were
found around 1385-68i MeV and 1419-22i MeV. The analysis of these data
also allowed to predict the cross section for the $K^- p \to \pi^0
\Sigma^0$ reaction in good agreement with data, without using the data
of that reaction in the fit. While in all known reactions the two poles
do not revert into two peaks, but only in different shapes 
\cite{Thomas:1973uh,Hemingway:1984pz,Niiyama:2008rt,prakhov,Moriya:2013eb,Moriya:2012zz,Moriya:2013hwg,Zychor:2007gf,fabbietti,Siebenson:2013rpa,Braun:1977wd},
the new data on electroproduction \cite{Lu:2013nza}, with two visible peaks around 1368
MeV and 1423 MeV, have been an unexpected surprise. The lower pole is
also broader than the higher one, as extracted from the photoproduction
data in \cite{Roca:2013av}.

  In the present paper we shall follow the strategy of \cite{Roca:2013av} and use only the photoproduction data of the  $\gamma p \to K^+ \pi^\pm \Sigma^\mp$ reactions measured in \cite{Moriya:2013eb} in order to extract the isospin I=1 amplitude in addition to the I=0 one extracted in \cite{Roca:2013av} from the 
$\gamma p \to K^+ \pi^0 \Sigma^0 $ data alone. This will allow us to shed light on the possible I=1 state around the $\bar K N$ threshold which has been often advocate. Indeed, in \cite{Oller:2000fj} hints of the existence of such state were discussed. The state became fuzzy in the analysis of \cite{cola}, showing up with sets of parameters with small SU(3) breaking, and reverting into a cusp with full SU(3) breaking. Some experimental support for such state has also been provided in \cite{Wu:2009nw,Gao:2010hy}. On the other hand, once again we will show that the data on photoproduction contain by themselves enough information to provide both the I=0 and I=1 amplitudes, by means of which, and without fitting the data, good results are obtained for the $\bar K N$ scattering amplitudes. The analysis conducted here will shows that, while no clear pole is found in the second Riemann sheet for the I=1 state (while a pole is found in another unphysical Riemann sheet), the amplitude in this channel is strong enough and has obvious repercussion in the photoproduction data, producing a clear split of the $\gamma p \to K^+ \pi^\pm \Sigma^\mp$ cross sections. Whether there is pole or not, the I=1 amplitude shows a clear enhancement close to the $\bar K N$ threshold and is visible as a pronounced cusp as a consequence of a strong attraction, in the border line of creating a quasibound bound  $\bar K N$ state.
We show that the situation is very similar to the one of the $a_0(980)$ resonance,
which is accepted as a resonance. The fact remains that whether one decides to call it
or not a resonance, the resonant like structure in the real axis has important repercussion in the photoproduction amplitudes and similarly can have relevant effects in many other observables.

\section{Unitarized meson-baryon amplitude}

The main ingredient of our analysis are the meson baryon amplitudes 
from the chiral unitary approach. 
There are many references where the details for the construction of the
meson-baryon unitarized amplitude can be found, (see for instance refs.~\cite{Oset:1998it,
Oller:2000fj,Hyodo:2006kg,Oset:2001cn}). It was also summarized in ref.~\cite{Roca:2013av}
for the $I=0$ case.
We next review it for the sake of completeness and including in addition
the evaluation of the $I=1$ channel.

The lowest order chiral Lagrangian for the interaction of the octet of 
Goldstone bosons with the octet of the low lying $1/2^+$ baryons
\cite{Bernard:1995dp}  provides the following 
tree level transition amplitudes in $s$-wave
\cite{Oset:2001cn}:

\ba
    V^I_{ij}(\rts)&=&-C^I_{ij}\frac{1}{4f^2}(2\rts-M_i-M_j)\nn\\
    &\times&\left(\frac{M_i+E_i}{2M_i}\right)^{1/2}
    \left(\frac{M_j+E_j}{2M_j}\right)^{1/2}, 
    \label{eq:WT}
\ea
where the superscript $I$ stands for the isospin, 
$\rts$ the center of mass energy, $f$ the averaged meson decay
constant $f = 1.123f_\pi$~\cite{Oset:2001cn} with $f_\pi = 92.4\mev$,
$E_i$ ($M_i$) the energies (masses) of the baryons of the $i$-th 
channel. The $C^0_{ij}$ coefficients, for isospin $I=0$, are given by 
\begin{equation} C^0_{ij} =\begin{pmatrix} 3 &
-\sqrt{\frac{3}{2}}  \\ -\sqrt{\frac{3}{2}}& 4 
\end{pmatrix} 
\label{eq:couplingI}. 
\end{equation}
The $i$ and $j$ subscripts represent the channels  $\bar K N$ and
$\pi\Sigma$ in isospin-basis. Note that we do not  consider  the other
possible channels in $I=0$,  $\eta\Lambda$ and $K\Xi$,  for the sake of
simplicity of the approach and because for the energies that we will
consider in this work the effect of those channels can be effectively
reabsorbed in the subtraction constants. 
The coefficients for isospin $I=1$ are
\begin{equation} C^1_{ij} =\begin{pmatrix} 3 & -1 & -\sqrt{\frac{3}{2}} \\
-1 & 2 &0 \\
-\sqrt{\frac{3}{2}}  & 0 & 0
\end{pmatrix} 
\label{eq:couplingI1}. 
\end{equation}
where the order of the channels are  $\bar K N$, 
$\pi\Sigma$ and $\pi\Lambda$.  We also neglect here the
$\eta\Sigma$ and $K\Xi$ 
states into the coupled channels equations
 since their thresholds are also very far from the energy region of interest 
 in the present work.

The chiral
unitary approach is based on
the implementation of 
unitarity of the scattering amplitude in coupled channels
and the exploitation of its
analytic properties. This is usually accomplished by means of
the Inverse Amplitude
Method \cite{dobado-pelaez,Oller:1998hw} or the N/D method 
\cite{Oller:1998zr,Oller:2000fj,Hyodo:2003qa}. In this latter
work the equivalence with the Bethe-Salpeter equation
used in \cite{Oller:1997ti} was established.
 Based on the $N/D$ method, the coupled-channel scattering amplitude 
$T_{ij}$ is given by the matrix equation
\begin{equation}
   T=[1-VG]^{-1}V ,
   \label{eq:BS}
\end{equation}
where $V_{ij}$ is the interaction kernel of Eq.~(\ref{eq:WT}) and the 
function $G_{i}$, or unitary bubble, is given by the dispersion integral of 
the two-body phase space 
$\rho_i(s)=2M_{i}q_i/(8\pi W)$, in a diagonal matrix 
form,  with $M_i$ the mass of the baryon of the meson baryon loop, $q_i$ the on shell momentum of the particles of the loop and $W$ the center of mass energy. 

This $G_i$ function is  
equivalent to the meson-baryon loop function
\begin{eqnarray}
G_{i} &=& i \, \int \frac{d^4 q}{(2 \pi)^4} \, \frac{M_i}{E_i
(\vec{q}\,)} \nn\\
&\times&\frac{1}{k^0 + p^0 - q^0 - E_i (\vec{q}\,) + i \epsilon} \,
\frac{1}{q^2 - m^2_i + i \epsilon} ~.
\label{gloop}
\end{eqnarray}
This integral is logarithmically divergent, 
and therefore it must be regularized, which
is usually carried out either with a three momentum cutoff
or with dimensional regularization in terms of a 
subtraction constant $a_i$. The connection and equivalence between
both methods was shown in Refs. \cite{Oller:1998hw,Oller:2000fj}.
In ref.~\cite{Oset:2001cn,cola}
 the values $a_{KN} = -1.84$, $a_{\pi\Sigma}=-2$
where used for the $I=0$ channels. In the present case, since we do not consider the 
$\eta\Lambda$ and $K\Xi$ channels, these subtraction constants may differ
slightly but we will allow to vary these constants in the
fit below.
For the $I=1$ channels, in ref.\cite{Hyodo:2011ur} the same value for 
$a_{KN}$, $a_{\pi\Sigma}$ as in the $I=0$ case was used and $a_{\pi\Lambda}=-1.83$ 
for the new channel in the $I=1$. 

\begin{figure}[!h]
\begin{center}
\includegraphics[width=0.9\linewidth]{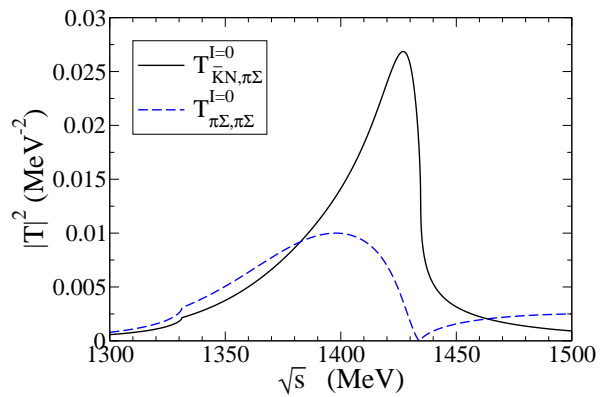}
\caption{(Color online) Modulus squared of the $I=0$ meson-baryon unitarized
amplitudes $T^{I=0}_{\bar K N,\pi\Sigma}$ (solid line) and 
$T^{I=0}_{\pi\Sigma,\pi\Sigma}$ (dashed line).}
\label{fig:t_MMMM}
\end{center}
\end{figure}

The amplitudes $T_{\bar K N\to\pi \Sigma}$ and $T_{\pi \Sigma\to
\pi \Sigma}$ for $I=0$ are depicted in fig.~\ref{fig:t_MMMM}. They
 produce two poles in the
second Riemann sheet of the complex energy plane at the positions 
$\sqrt{s_0}=1387-67i\mev$,
and $1437-13i\mev$. Note that the poles come dynamically from the
non-linear dynamics involved in the implementation of unitarity in the
meson-baryon scattering amplitude, without
the need to include the poles as explicit degrees of freedom. This is
what is usually called {\it dynamically generated} resonance or
meson-baryon molecule.
It is worth mentioning that the unitarized amplitudes
provide the
actual meson-baryon scattering amplitudes, not only the poles of the
resonance in the complex plane. Indeed the resonant shapes
of the amplitudes around the 1400~MeV region are far from looking
like Breit-Wigner shapes. Therefore fits to experimental
data assuming 
Breit-Wigner resonant shapes
are not suitable for this resonance and a model like the present
one, in the line of
implementing unitarity in coupled channels, is called for in order to
reproduce or fit experimental data where these amplitudes
are relevant.
\begin{figure}[!h]
\begin{center}
\includegraphics[width=0.9\linewidth]{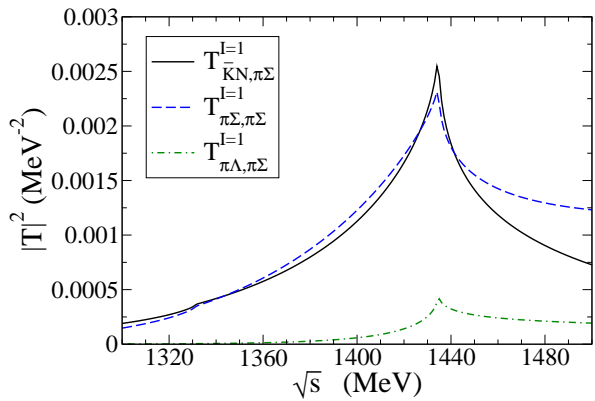}
\caption{(Color online) Modulus squared of the $I=1$ meson-baryon unitarized
amplitudes $T^{I=1}_{\pi\Sigma,\pi\Sigma}$ (solid line), 
$T^{I=1}_{\bar K N,\pi\Sigma}$ (dashed line) and 
$T^{I=1}_{\pi\Lambda,\pi\Sigma}$ (dashed-dotted
line) .}
\label{fig:t_MMMM1}
\end{center}
\end{figure}
In fig.~\ref{fig:t_MMMM1} we show the amplitudes
 $T_{\bar K N\to\pi \Sigma}$, $T_{\pi\Sigma\to\pi \Sigma}$
and $T_{\pi\Sigma\to\pi \Lambda}$ in $I=1$.
 In this case there is no pole associated to the visible increase of
strength appreciable at threshold in the amplitudes. 
We will elaborate further on this issue later on and we will
 discuss on the possible connexion to an
actual $I=1$ resonance in the next section.

\section{Fit to photoproduction data}

In our previous work \cite{Roca:2013av} we only considered the $\gamma p\to K^+
\pi^0\Sigma^0$ data of \cite{Moriya:2013eb} since this particular reaction filters the
$I=0$ and therefore these were the only data used in \cite{Roca:2013av}.
However, in the present work we are also
interested in the $I=1$ channel in order to try to make conclusions from a
possible resonance in $I=1$.
If one looks at the isospin decomposition of the final $\pi\Sigma$ states,
\ba
|\pi^0\Sigma^0\rangle&=&\sqrt{\frac{2}{3}}|2 \,0\rangle-\frac{1}{\sqrt{3}}|0\,
0\rangle,  \nn\\
|\pi^+\Sigma^-\rangle&=&-\frac{1}{\sqrt{6}}|2 \,0\rangle-\frac{1}{\sqrt{2}}|1\,
0\rangle -   \frac{1}{\sqrt{3}}|0\,0\rangle,       \nn\\
|\pi^-\Sigma^+\rangle&=&-\frac{1}{\sqrt{6}}|2 \,0\rangle+\frac{1}{\sqrt{2}}|1\,
0\rangle -   \frac{1}{\sqrt{3}}|0\,0\rangle,      
\ea
it is clear then that one must also include 
the  $\gamma p\to K^+ \pi^+\Sigma^-$, $\gamma
p\to K^+ \pi^-\Sigma^+$ in the analysis.\footnote{We neglect the isospin $I=2$ since it
is very small and non-resonant in the energy region of interest in the present work.}

The main observable measured for this reaction is 
the $\pi\Sigma$ invariant mass distribution for the different allowed charge
combinations (see fig.~\ref{fig:fitVcte} below).

\begin{figure}[!h]
\begin{center}
\includegraphics[width=0.9\linewidth]{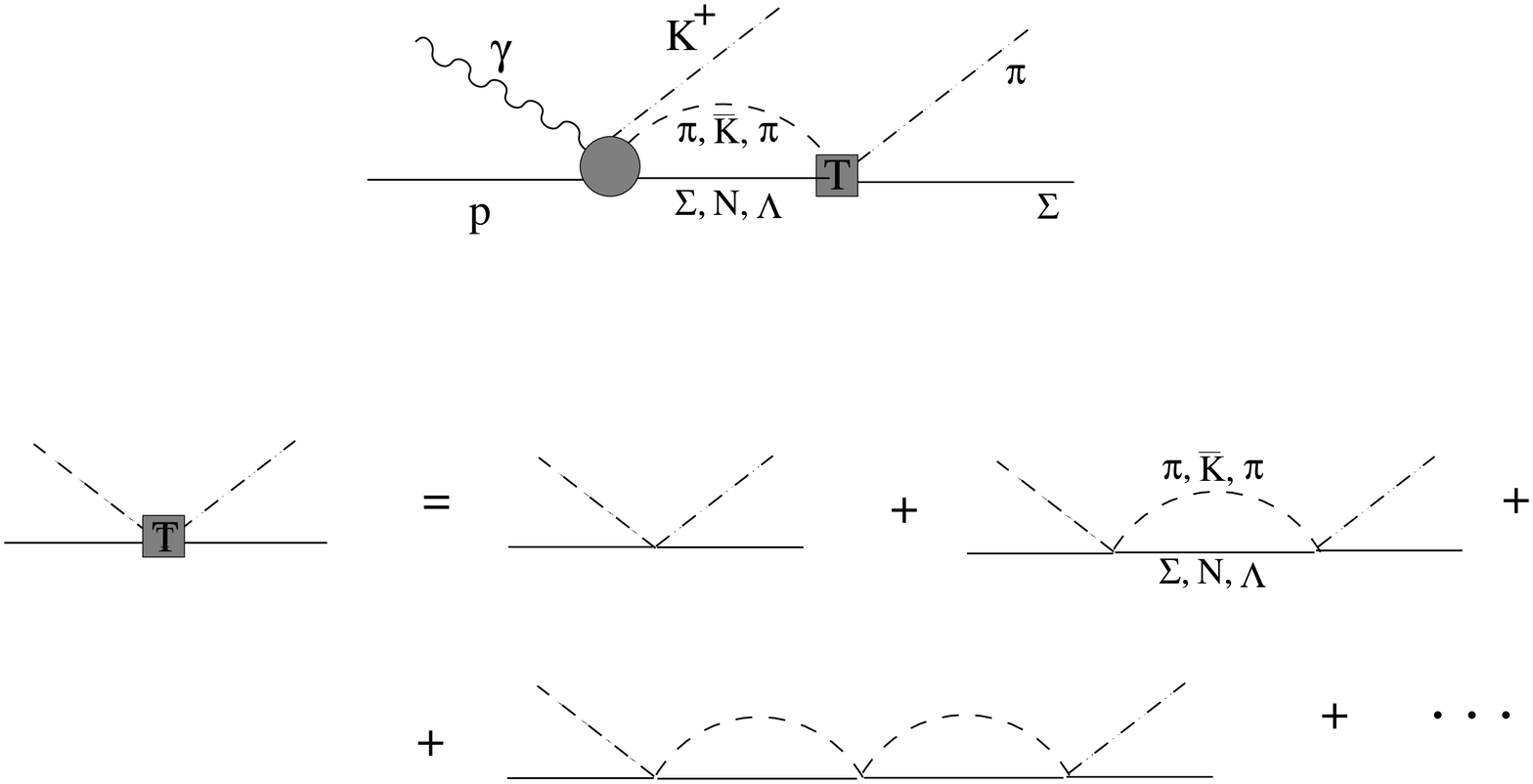}
\caption{General mechanisms for the photoproduction amplitudes}
\label{fig:diagrams}
\end{center}
\end{figure}

Since the $\Lambda(1405)$ is  dynamically
generated from the final state interaction of the meson-baryon pair
produced, and we also seek for a possible generated resonance from the meson-baryon
scattering in $I=1$,
 the most general mechanisms for the
photoproduction reaction are those depicted in fig.~\ref{fig:diagrams}.
The photoproduction can proceed by the production of either a
$\pi\Sigma$ or $\bar K N$ pair for $I=0$ and $I=1$ and also by $\pi \Lambda$ for the
$I=1$ case. This initial production is represented by the thick circle in  fig.~\ref{fig:diagrams}.
The initial meson-baryon pair then rescatters to produce the final $\pi\Sigma$, accounted for by the
unitarized scattering amplitude explained in the previous section.
Note that a possible contact mechanism of direct $\pi\Sigma$ production
would contribute to the background and we do not consider it 
since a proper background subtraction has been done in the 
experimental analysis.

Based on fig.~\ref{fig:diagrams} 
it is immediate to realize that the
amplitudes for the photoproduction process can be generally
written as
\ba
t_{\gamma p\to K^+ \pi^0\Sigma^0}(W)&=&b_0(W) G^{I=0}_{\pi\Sigma} T^{I=0}_{\pi\Sigma,\pi\Sigma}\nn\\
&+&c_0(W) G_{\bar K N}^{I=0} T^{I=0}_{\bar K N,\pi\Sigma}\ , \nn\\
t_{\gamma p\to K^+ \pi^\pm\Sigma^\mp}(W)&=&b_0(W) G^{I=0}_{\pi\Sigma} T^{I=0}_{\pi\Sigma,\pi\Sigma}\nn\\
&+&c_0(W) G_{\bar K N}^{I=0} T^{I=0}_{\bar K N,\pi\Sigma} \nn\\
&\pm&\sqrt{\frac{3}{2}} \bigg( 
    b_1(W) G^{I=1}_{\pi\Sigma} T^{I=1}_{\pi\Sigma,\pi\Sigma} \nn \\
   &+& c_1(W) G^{I=1}_{\bar K N} T^{I=1}_{\bar K N,\pi\Sigma} \nn\\
   &+& d_1(W) G^{I=1}_{\pi\Lambda} T^{I=1}_{\pi\Lambda,\pi\Sigma}
			    \bigg)
\label{eq:tampli}
\ea
with $W$ the energy of the $\gamma p$ interaction. 
The subindex in the $b$, $c$ and $d$ coefficients stand for the isospin.
Note that the only difference between the 
$\gamma p\to K^+ \pi^+\Sigma^-$ and the $\gamma p\to K^+ \pi^-\Sigma^+$ amplitudes is the sign of the
$I=1$ contributions.
The coefficients $b$, $c$ and $d$ may in general depend on  $W$ 
and hence we consider 9 sets of them in order to
account for the 9 different energies $W$ provided
by the experimental result
of CLAS \cite{Moriya:2013eb}. On the other hand the relative weight between the 
 different $GT$ addends may be complex in
 general,
 therefore we allow  the $b_1$, $c_0$, 
$c_1$ and $d_1$  to be complex and keep $b_0$
real since a global phase in the total amplitude is irrelevant. We will refer the $b$, $c$ and $d$
coefficients by {\it initial production} (IP) parameters  in the following.

Note that, as in ref.~\cite{Roca:2013av}, we try to keep the analysis as model independent as
possible in order to ease its implementation by experimentalist groups. Therefore we
intentionally avoid proposing
any model for the initial
photoproduction mechanisms (filled circle in Fig.~\ref{fig:diagrams}) which are
effectively encoded in the IP parameters.
In the actual reaction they would contain
a rich dynamics that could count for some
contribution from $N^*$ resonances, crossed diagrams, 
$t$-channel processes,
etc., projected over $s$-wave. Since we take the coefficients energy dependent,
 $b(W)$, $c(W)$,
$d(W)$
 the
fit to the data can accommodate this dynamics without explicitly taking
it into account.
 
 Since we are fitting 9 different energies,  we have
in total 81 IP parameters. We are aware that this figure may look large but none of these coefficients 
affect the meson-baryon
scattering amplitude where the resonant dynamics actually stems from.
 One has to view the fit from the perspective that the data for one
energy will provide the small subset of IP parameters for that particular
energy. Only the parameters of the potential, that we 
will consider and explain later in this paper,
affect all the data. This
situation is similar  to the fit conducted to pionic atoms to extract
  neutron radii in \cite{neutronrad}. In that problem there were 19
  parameters for 19 neutron radii and 6 parameters for the potential.
  Again, each of these 19 parameters affected only the data on shifts
  and widths of a single pionic atom and the 6 parameters of the
  potential affected all the data. The fits worked without problems and
  the set of neutron radii obtained is considered nowadays the most
  valuable experimental source of neutron radii, together with the
  information obtained from antiprotonic atoms in \cite{pabarneutron}. 

We first  fit the IP parameters to the photoproduction $\pi\Sigma$
invariant mass distribution data
using for
the unitarized amplitudes the expression and parameters explained in the
previous section.
Note that in this first step only the photoproduction vertex is allowed
to vary and the chiral unitary approach amplitudes are fixed.

\begin{figure*}[h]
\begin{center}
\includegraphics[width=0.99\textwidth]{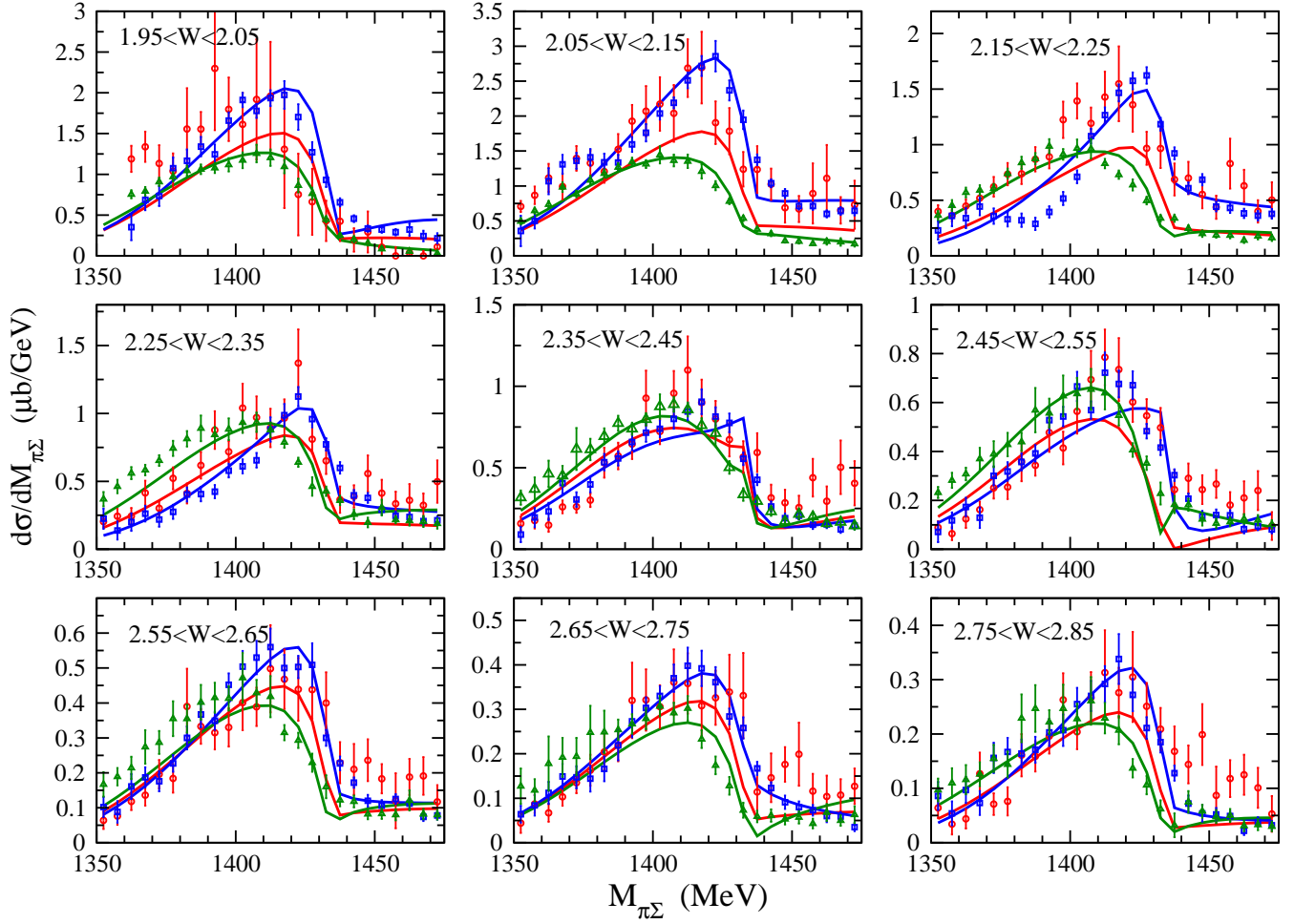}
\caption{(Color online) Fit to photoproduction data with fix unitary amplitudes, 
$\alpha_i=1$, $\beta_i=1$. Red: $\pi^0
\Sigma^0$; blue: $\pi^-
\Sigma^+$, green: $\pi^+\Sigma^-$. Experimental data from ref.~\cite{Moriya:2013eb}.}
\label{fig:fitVcte}
\end{center}
\end{figure*}

In the evaluation of the theoretical invariant mass distribution the three body phase
space has been averaged within the experimental $W$ bin, $[W-0.05,W+0.05]\gev$, for
every $W$. We perform the fit in the range $M_{\pi\Sigma}\in[1350,1475]\mev$.
The result of this fit is shown in Fig.~\ref{fig:fitVcte}.
One can see that the fit is visually fair for most of the energies,
which means that the actual meson-baryon amplitudes must not be much far
from those predicted by the chiral unitary approach. However a better $\chi^2/dof$ that
the one obtained in this fit  ($\chi^2/dof=4.6$)  would be desirable.

It is worth stressing again that what we actually want in the present work is not to
calculate what the chiral unitary approach predicts for the poles of the
$\Lambda(1405)$ or a possible $I=1$ resonance, but to extract them from the
experimental photoproduction  data. Therefore we can try to get results with better
$\chi^2/dof$ by allowing the basic chiral unitary model to vary slightly.  In this way
we could obtain a fine tuning of the chiral unitary model and then of the position of
the $\Lambda(1405)$ poles and try to see if some $I=1$ resonance shows up. In order to
do this we multiply each coefficient of the  potentials of the unitary amplitudes,  
Eqs.~(\ref{eq:couplingI}) and (\ref{eq:couplingI1}), by one real parameter $\alpha_i$
and  hence the new coefficient matrices that we consider now are given by
 \begin{equation} C^0_{ij} =\begin{pmatrix} 3 \alpha^0_{11} &
-\sqrt{\frac{3}{2}} \alpha^0_{12}  \\ -\sqrt{\frac{3}{2}}\alpha^0_{12}& 4 \alpha^0_{22}
\end{pmatrix} 
\label{eq:couplingIalpha}
\end{equation}
for isospin $I=0$
and
\begin{equation} C^1_{ij} =\begin{pmatrix} 3 \alpha^1_{11}& -\alpha^1_{12} 
& -\sqrt{\frac{3}{2}}\alpha^1_{13} \\
- \alpha^1_{12} & 2\alpha^1_{22} &0 \\
-\sqrt{\frac{3}{2}}  \alpha^1_{13}& 0 & 0
\end{pmatrix} 
\label{eq:couplingI1alpha}
\end{equation}
for isospin $I=1$.

Furthermore we also allow to vary the subtraction constants from
the regularization of the loop functions by multiplying each of them by 
a free parameter, $\beta_i$:
   $a_{KN}\to \beta_1 a_{KN}$, $a_{\pi\Sigma}\to \beta_2
a_{\pi\Sigma}$ and $a_{\pi\Lambda}\to \beta_3
a_{\pi\Lambda}$.
We  will refer to the $\alpha$ and $\beta$ parameters by {\it potential} parameters in the following
(even though the $\beta$ coefficients do not affect the potential, but we do this 
 just to ease the
nomenclature).
Therefore, the chiral unitary amplitudes depend on 10 free parameters to be fitted,
$\alpha_i$, $\beta_i$, but only 5 of them affect the $I=0$ amplitude and 7 the $I=1$.
With the potential obtained from the fit we shall search
 for the positions of the two $\Lambda(1405)$ poles and look for a possible $I=1$ resonance in the
 range of energy considered.

If at this point we carry on a global fit allowing for all the parameters to be  free
from the beginning in the fitting algorithm, there are many local minima of the 
$\chi^2$  function, most of them having clearly unphysical values of the parameters.
Therefore it is very difficult to get and identify  an absolute minimum. Actually many 
minima have $\chi^2$ very similar but with very different values of the parameters,
which spoils the statistical significance of the fit and the possible physical
conclusions. In order to get physically meaningful results, we implement the following
strategy in the line of the one used in ref.~\cite{Roca:2013av}: As mentioned above, the
previous fit of fig.~\ref{fig:fitVcte}, i.e. fixing the potential parameters to 1, is
already reasonably fair,  and the potential is consistent with data of scattering
\cite{Oset:1998it},  hence a good physical global fit should not be very far from
having values of $\alpha_i\sim 1$, $\beta_i\sim 1$. Therefore, in a first step, we
start from the fit of fig.~\ref{fig:fitVcte}, which was obtained fixing the potential
parameters to 1 ($\alpha_i=1$, $\beta_i=1$), but fixing now the IP parameters and
allowing only the potential parameters to change.  In a next step, we fix the new
potential parameters obtained in the previous step and fit again the IP parameters. We
iterate the process alternating between fitting the IP or fitting the potential
parameters until we get a convergence of the value of the $\chi^2$.  In this way we
obtain a minimum of the $\chi^2$ with potential parameters not very different from 1
which are then physically meaningful.

\begin{figure*}[h]
\begin{center}
\includegraphics[width=0.99\textwidth]{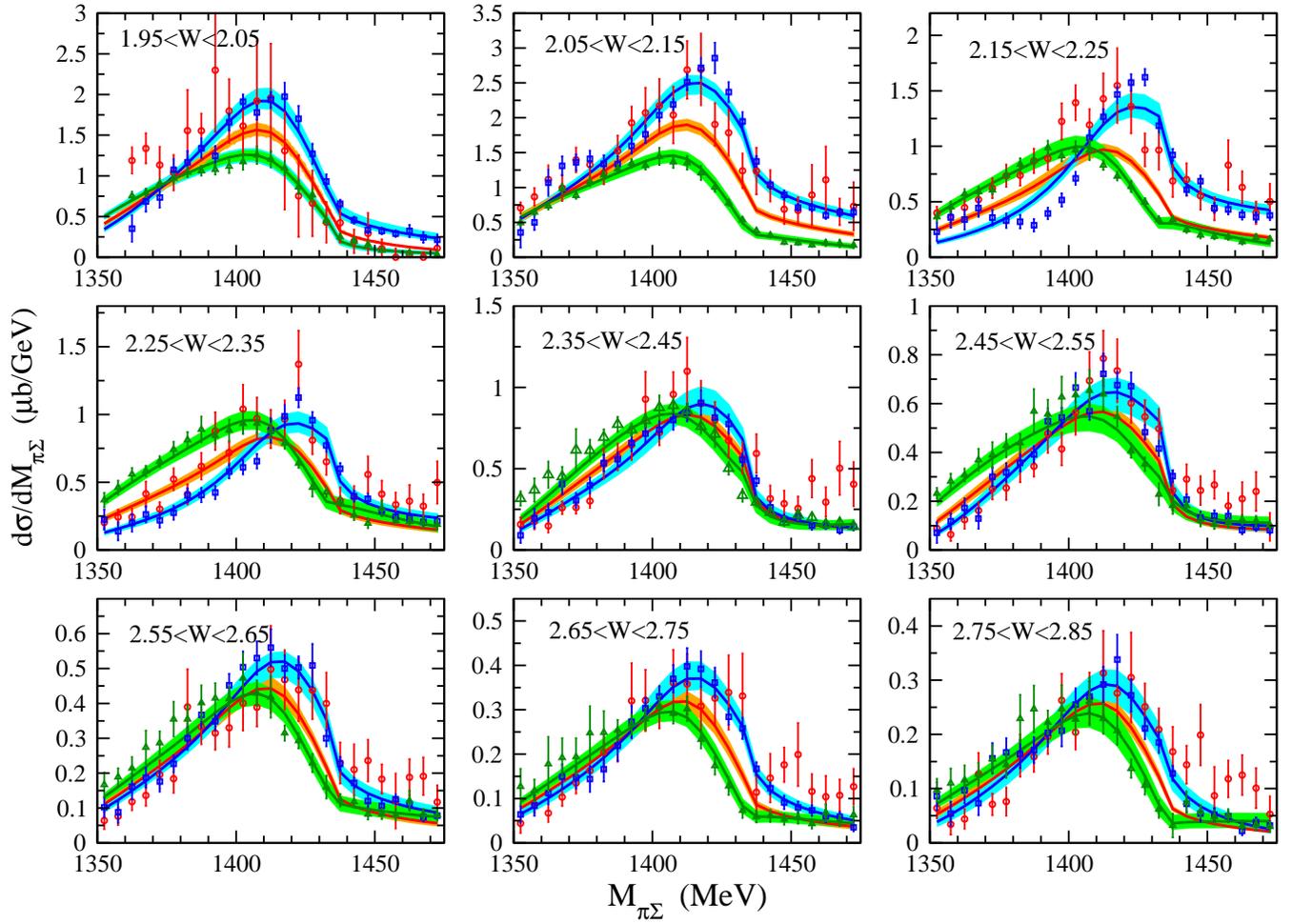}
\caption{(Color online) solution from the fit procedure described in the text}
\label{fig:fitsols12}
\end{center}
\end{figure*}

After this iterative procedure we get the result shown in fig.~\ref{fig:fitsols12},
which has  $\chi^2/dof=2.1$. The bands account for the uncertainties of the fit at one
standard deviation confidence level. The potential parameters obtained are shown in
table~\ref{tab:resultsalpha}.

\begin{table}[h]
\caption{Parameters of the unitarized amplitudes}
\begin{center}
\begin{tabular}{c c c c c c c c c c} 
\hline\hline 
 $\alpha^0_{11}$ & $\alpha^0_{12}$  &  $\alpha^0_{22}$    & $\alpha^1_{11}$ & $\alpha^1_{12}$  &  $\alpha^1_{13}$    &   $\alpha^1_{22}$ &$\beta_1$
 &$\beta_2$&$\beta_3$ \\ \hline
 $1.037$ & $1.466$ &  $ 1.668$  &   
      $ 0.85$     &
  $ 0.93$ & $ 1.056$ & $ 0.77$ & $1.187$ & $ 0.722$  & $1.119$\\ \hline\hline
 \end{tabular}
\end{center}
\label{tab:resultsalpha}
\end{table}

It is important to note that the parameters obtained 
are not very different from 
one. This means that allowing for just a small variation in the 
parameters of the chiral unitary approach the photoproduction data can
be nicely reproduced.

In table~\ref{tab:resultspoles} we show the results obtained  for the pole positions
in the complex energy ($\sqrt{s}$) plane in unphysical Riemann sheets of the scattering
amplitudes. 
\begin{table}[h]
\caption{Pole positions (in MeV) in the complex energy plane of the scattering
 amplitudes and modulus of the couplings to the different channels.}
\begin{center}
\begin{tabular}{|c |c| c| c|} 
\hline\hline  &
\multicolumn{2}{c|}{$I=0$} & $I=1$     \\ \hline
poles          &  $1352-48i$ & $1419-29i$ &  $-$ \\ \hline
$|g_{\bar K N}|$ &  $2.71$     &  $3.06$    &  $-$ \\  \hline
$|g_{\pi\Sigma}|$&  $2.96$     &  $1.96$    &  $-$ \\ \hline\hline
 \end{tabular}
\end{center}
\label{tab:resultspoles}
\end{table}
In the table we also show the modulus of the couplings to the different isospin
meson-baryon channels obtained from the residues  of the unitarized meson-baryon
scattering amplitudes at the pole positions, since close to the pole position the
amplitude can be approximated by its Laurent expansion where the dominant term is given
by

\be
    T_{ij}(\sqrt{s})
    =
    \frac{g_i\,g_j}{\sqrt{s}-\sqrt{s_\textrm{pole}}},
    \label{eq:Tpole}
\ee
for an $s$-wave resonance, where the position of the pole can be identified with the
 mass, $M_R$, and
width, $\Gamma_R$, of the resonance by
$\sqrt{s_\textrm{pole}}=M_R-i\Gamma_R/2$ for a pole not very far from the real axis.
 Consequently the residue of $T_{ij}$ at the pole 
position gives $g_ig_j$, where $g_i$ is the effective coupling of the 
dynamically generated resonance to the $i$-th channel.

\begin{figure}[!h]
\begin{center}
\includegraphics[width=0.9\linewidth]{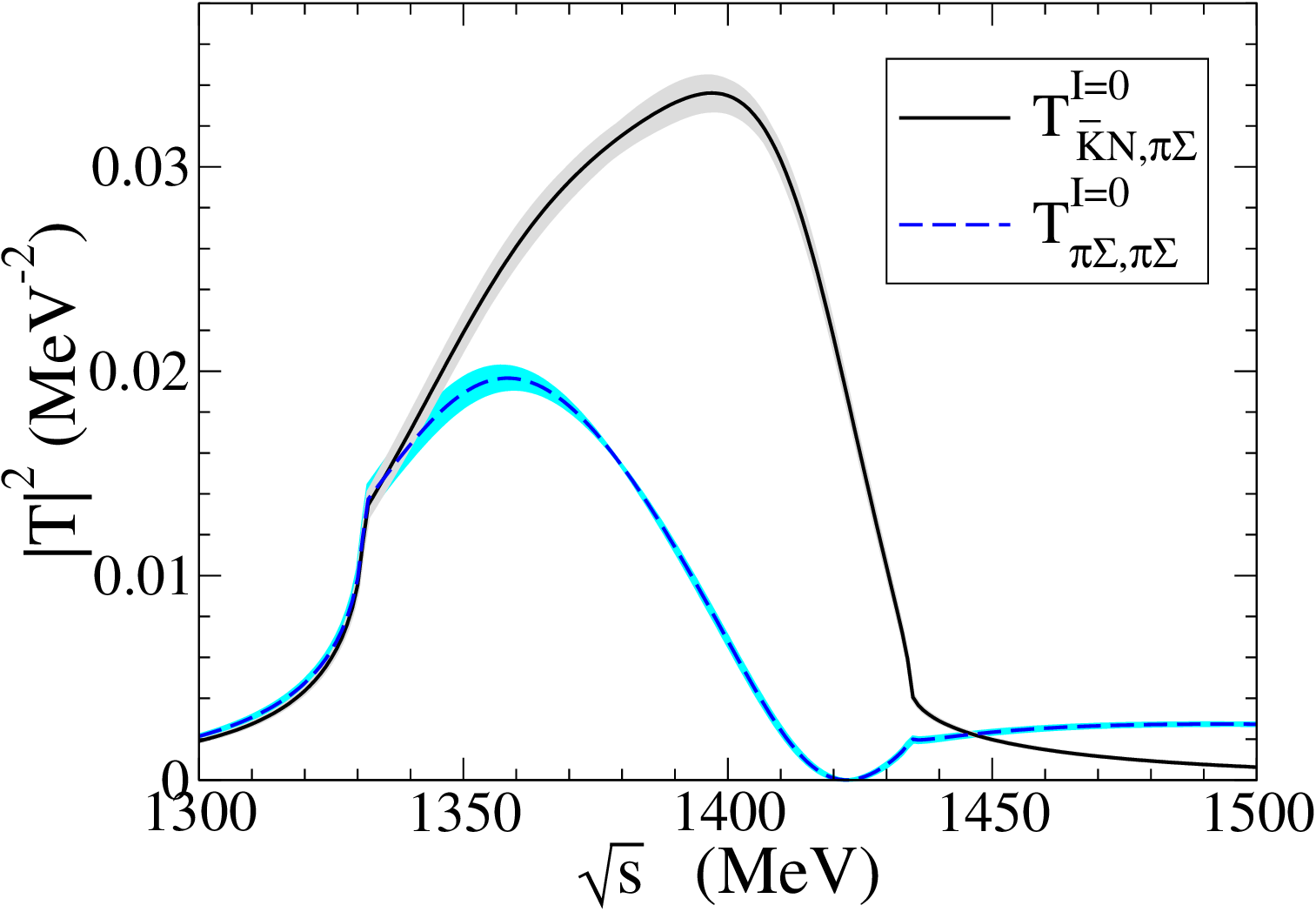}
\caption{(Color online) Modulus squared of the $I=0$ meson-baryon unitarized
amplitudes $T^{I=0}_{\pi\Sigma,\pi\Sigma}$ (solid line), 
$T^{I=1}_{\bar K N,\pi\Sigma}$ (dashed line).}
\label{fig:t_MMMM0fit}

\end{center}
\end{figure}

\begin{figure}[!h]
\begin{center}
\includegraphics[width=0.9\linewidth]{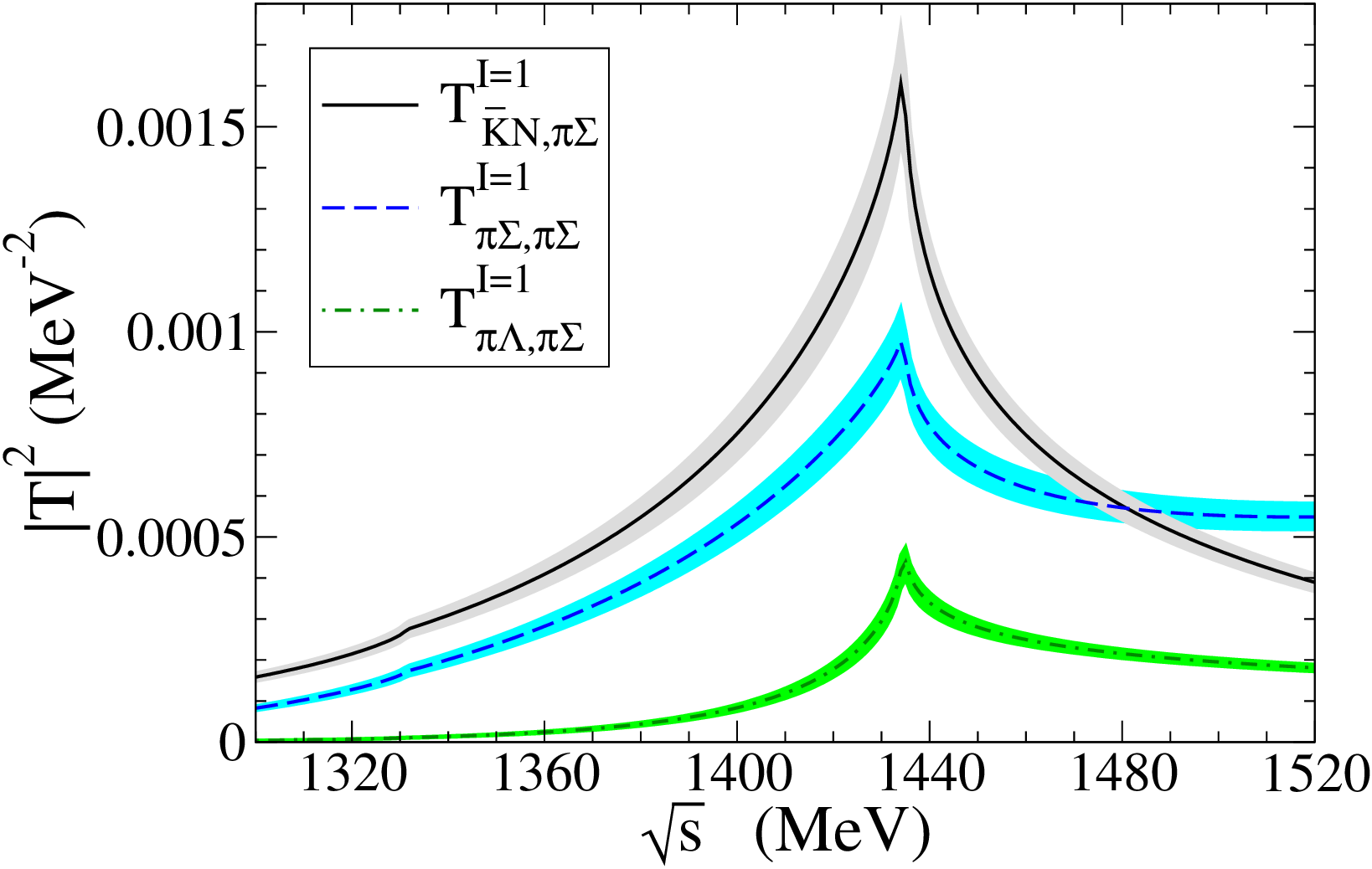}
\caption{(Color online) Modulus squared of the $I=1$ meson-baryon unitarized
amplitudes $T^{I=1}_{\pi\Sigma,\pi\Sigma}$ (solid line), 
$T^{I=1}_{\bar K N,\pi\Sigma}$ (dashed line) and 
$T^{I=1}_{\pi\Lambda,\pi\Sigma}$ (dashed-dotted
line) .}
\label{fig:t_MMMM1fit}
\end{center}
\end{figure}

The poles have been looked for in the usual unphysical
Riemann sheet of the scattering amplitudes which is
defined in the following way:
The analytic 
structure of the scattering amplitude is determined by the loop function 
$G_i$ (Eq.~\ref{gloop}). The $G_i$ function in the second Riemann sheet (RII) can be obtained
from the one in the first sheet (RI) by \cite{Roca:2005nm}
\begin{equation}
G_i^{II}(\sqrt{s})=G_i^{I}(\sqrt{s})+iM_i\frac{q_i}{2\pi \sqrt{s}},
\label{eq:GII}
\end{equation}
\noindent
with $q_i$ the center of mass meson or baryon momentum
with $\im(q_i)>0$. When looking for poles we use $G_j^I(\sqrt{s})$ for 
$\re(\sqrt{s})<m_j+M_j$ and $G_j^{II}(\sqrt{s})$ for $\re(\sqrt{s})>m_j+M_j$.
 This prescription, which we
will refer to as usual unphysical Riemann sheet in the following, 
gives the pole positions closer to those of the corresponding Breit-Wigner 
forms on the real axis.
In this way, no pole is found for $I=1$ but there is a pole located at $1522-14i\mev$
in another unphysical Riemann sheet
defined by going to RII for $\pi\Lambda$ and $\pi\Sigma$ channels but not for  $\bar K N$ despite
being above the $\bar K N$ threshold ($1435\mev$).
This pole in this different Riemann sheet does not produce a Breit-Wigner shape in the real axis
in the physical sheet
but makes the shape of the amplitude below the $\bar K N$ threshold to follow the shape of the tail
of
that pole but then decrease above threshold with a non resonant shape. This means that, for the $I=1$ amplitudes considered here, even if there is not an explicit pole in the usual unphysical Riemann
sheet, an accumulation of strength is present on the real axis in the physical sheet, under the
appearance of a cusp.
In
Figs.~\ref{fig:t_MMMM0fit} and \ref{fig:t_MMMM1fit} we show the $I=0$ and $I=1$ meson-baryon
amplitudes with the result of the fit.
In Fig.~\ref{fig:t_MMMM1fit}, $I=1$ case, one can see the aforementioned increase
 of strength and cusp aspect
at the $\bar K N$ threshold which could be perceived as a resonance in an actual experiment.
In order to see that the $I=1$ amplitude still has to do 
with a resonant structure, in spite the fact that it does not have a pole in 
the usual Riemann sheet
described above, let us do a mathematical play consisting
on varying by hand some of the potential parameter:
\begin{figure}[!h]
\begin{center}
\includegraphics[width=0.9\linewidth]{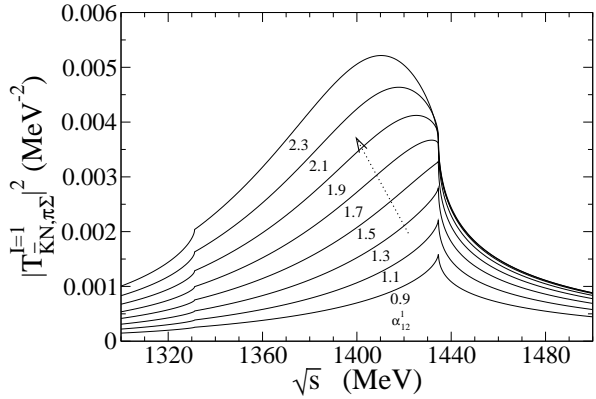}
\caption{Evolution of the $I=1$ $\bar K N \to \pi\Sigma$ scattering amplitude
($T^{I=1}_{\bar K N,\pi\Sigma}$) as a function of the $\alpha^1_{12}$ coefficient.}
\label{fig:t_MMMM1_running}
\end{center}
\end{figure}
\begin{figure}[!h]
\begin{center}
\includegraphics[width=0.9\linewidth]{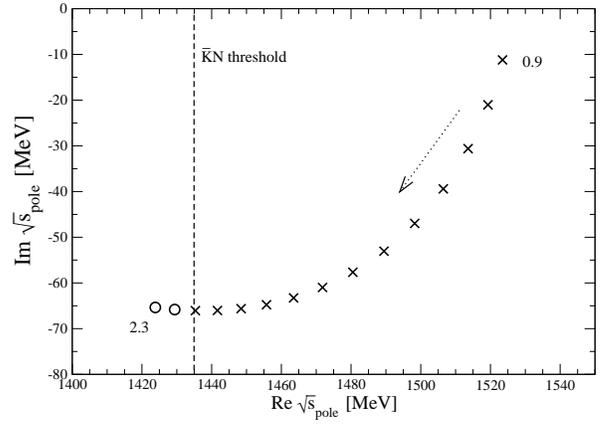}
\caption{Evolution of the $I=1$ pole position as a function of the $\alpha^1_{12}$ coefficient.}
\label{fig:pole_running}
\end{center}
\end{figure}

Let us change  by hand the $\alpha^1_{12}$ parameter, for example, from 0.9 
(close to the result of the fit)
to 2.3. The resulting $T^{I=1}_{\bar K N,\pi\Sigma}$ amplitude
and the position of the poles found in different sheets are shown 
 in Figs.~\ref{fig:t_MMMM1_running} and \ref{fig:pole_running}. All the poles are found in Riemann
 sheets for which RI is used for the loop of the $\bar K N$ channel. This means that below 
 $\bar K N$ threshold (poles represented by circles Fig.~\ref{fig:pole_running})
 the Riemann sheet is the usual unphysical one described above but that 
 is not the case for poles whose real part is located above
 threshold (poles represented by crosses). We can see
 that the shape of the resonance gets distorted
in a continuous way as we change the parameter.
 However for a short range of the values of the parameter considered,
there is a pole in the usual unphysical sheet
which eventually disappears from that sheet, when the real part crosses the $\bar K N$ threshold. 
The poles depicted
above threshold are in the other unphysical sheet described above. That means that,
in spite the fact that there is no pole in the usual unphysical sheet 
for the particular value of the parameters obtained
 in the fit, the amplitude is continuously connected with a situation where there is a usual 
resonance pole and then
 somehow the amplitude is aware and reflects the existence of the pole for a nearby value
  of the parameter.
This is not a strange case since an analogous situation also shows up for the $a_0(980)$
resonance in the pseudoscalar-pseudoscalar scattering in the scalar isovector channel. In that case
there is a pole
very close to the $K \bar K$ threshold and for small variations in the parameters of the potential
the pole disappears from the  usual unphysical sheet and goes above threshold to the sheet where the
loop function for $K\bar K$ is RI. In spite of this fact, 
everybody considers the $a_0(980)$ as a resonance \cite{Oller:1998hw}.

\begin{figure}[h]
\begin{center}
\includegraphics[width=0.95\linewidth]{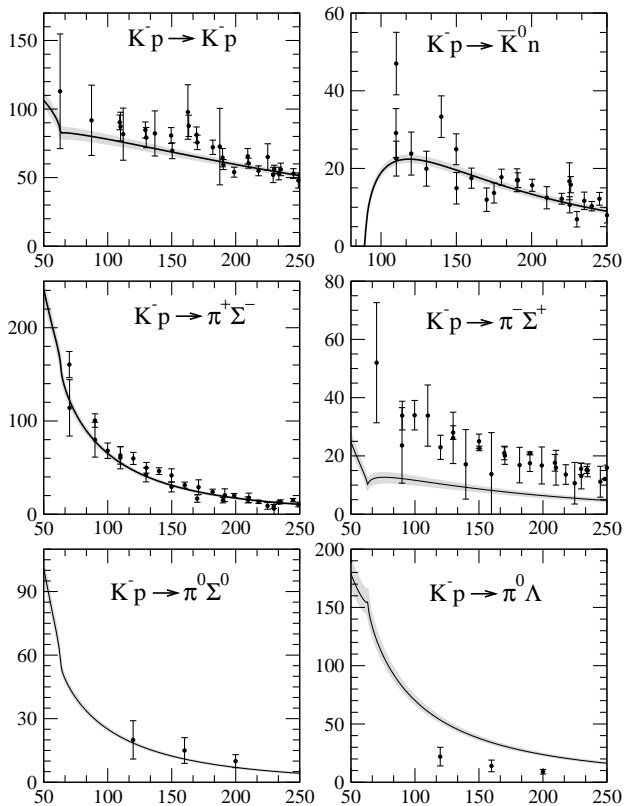}
\caption{Predicted $K^-p$ cross sections (in mb). Experimental
data from ref.~\cite{cross_sections_kp_exp}.}
\label{fig:crosskp}
\end{center}
\end{figure}

In order to make further checks that the fit obtained is physically
acceptable, 
we  calculate now the cross section for $K^-
p\to MB$ for the meson-baryon final channels $K^-p$, $K^0n$, $\pi^+\Sigma^-$, $\pi^-\Sigma^+$, $\pi^0\Sigma^0$ and $\pi^0\Lambda$.
The results are shown in Fig.~\ref{fig:crosskp} in comparison to experimental data 
\cite{cross_sections_kp_exp}.
Note that the  results in Fig.~\ref{fig:crosskp} are genuine non-trivial predictions
since 
the fit is only done to the
photoproduction data. The agreement of the 
$K^-p$ cross sections to experimental data is remarkable.

Another experimental data usually considered in other theoretical works
\cite{ollernew,mai,Ikeda:2011pi,Ikeda:2012au}
regarding the $\Lambda(1405)$
resonance are the energy shift and width of the kaonic hydrogen in the $1s$ 
state from the SIDDHARTA experiment at
DAFNE \cite{Bazzi:2011zj}, which are reported to be $\Delta
E-i\Gamma/2=(283\pm42)-i(271\pm55)\textrm{ eV}$. This value is related to the $K^-p$ scattering length
and therefore to the $K^-p\to K^-p$ amplitude at threshold. (For explicit mathematical expressions see
refs.~\cite{ollernew,mai,Ikeda:2011pi,Ikeda:2012au}).
With the values of the parameters in table~\ref{tab:resultsalpha} we obtain  $\Delta
E-i\Gamma/2=(194\pm 4)-i(301\pm 9)\textrm{ eV}$, which compares reasonably well to
 the experimental SIDDHARTA
data.

On the other hand, in a different fit to the CLAS data made by some members of that
collaboration \cite{Schumacher:2013vma},
two different kind of fits were performed: one only to the $\pi^0 \Sigma^0$ data, to which only the
$I=0$ channel contributes and one to all the photoproduction data. 
The amplitudes in that fit are parametrized as 
(Eq.(5) of \cite{Schumacher:2013vma})
\be
t_I(m)=C_I(W)e^{i\Delta\phi_I}B_I(m),
\ee
where $C_I(W)$ is a weight factor, $\Delta\phi_I$ a phase and $B_I(m)$ a
Breit-Wigner function. As one can see, the weight is allowed to depend
on the photon energy, $W$, but not its phase. But even more restrictive is
the fact that the shapes of the resonances, $B_I(m)$, are Breit-Wigner 
shapes and chosen independent of the
photon energy. This neglects the possibility that the amplitudes
 $\gamma p \to K^+ \pi
\Sigma$ are superpositions of the amplitudes corresponding to the different
poles with relative weights that depend on the photon energy.
Furthermore, as seen in the plots of the amplitudes throughout the present work, the resonant
amplitudes are far from being Breit-Wigner like.

With the fit to only the $I=0$ part introducing two poles,
 the authors in \cite{Schumacher:2013vma} get
for the mass and width of the resonances (all in MeV) $M=1329$, $\Gamma=20$, for one of the $I=0$ resonances
and $M=1390$, $\Gamma=174$ for the other one. This has to be compared to the fit only to $I=0$ that we did 
in ref.~\cite{Roca:2013av}; $M=1368$, $\Gamma=108$, and $M=1416$,
 $\Gamma=48$~\footnote{Note that there is a slight difference between these values and
 those reported in ref.~\cite{Roca:2013av}. This is due to a change in some of the
CLAS experimental data from those reported in \cite{Moriya:2012zz} to those in 
\cite{Moriya:2013eb}.}.
The differences are due to the reasons explained above and in ref.~\cite{Roca:2013av}.
The other fit performed in ref.~\cite{Schumacher:2013vma} includes two Breit-Wigners for 
$I=0$ and one for $I=1$ and they get 
$M=1338$, $\Gamma=44$, for one of the $I=0$ resonances
and $M=1384$, $\Gamma=76$, for the other one and $M=1357$, $\Gamma=54$, for the $I=1$. This fit should be
compared to ours in the present paper, (see table~\ref{tab:resultspoles}).
The difference in the results between our fit and CLAS' is understandable considering the caveats
explained above.

\section{Summary} 

We have implemented an strategy to obtain information on the $I=0$ and $I=1$ meson-baryon
scattering amplitudes from  $\gamma p \to K^+ \pi \Sigma$ experimental data. 
The idea is based on leaving the photoproduction vertices as model independent as possible, to ease
the implementation by experimental groups, 
parametrizing them by coefficients dependent on energy to be fitted to the photoproduction data.
The resonant structures come from the meson-baryon final state interaction implemented by 
amplitudes
inspired by the
chiral unitary approach but slightly modified  with free coefficients. These coefficients and those of
the linear combinations were fitted to the data and a good solution is obtained.
We provide the position of the two $\Lambda(1405)$ poles (predicted by the chiral
unitary approach) and
we have discussed the possible existence of an $I=1$ resonance around the $\bar K N$ threshold. In
spite the fact that there is not a pole in the usual unphysical Riemann sheet connected to the
physical one, we have discussed that there is a resonant structure in $I=1$ around
to the $\bar K N$ threshold.

Once the solution of the fit is established, we have obtained
fair results for the cross sections of the $K^-
p\to MB$ for the meson-baryon final channels $K^-p$, $K^0n$,
 $\pi^+\Sigma^-$, $\pi^-\Sigma^+$, $\pi^0\Sigma^0$ and $\pi^0\Lambda$ and for experimental data on
 kaonic hydrogen.

In the analysis carried out in the present
work we show that the information encoded in the photoproduction data is valuable to obtain the
information on the resonant content of the $I=0$ and $I=1$ channels in the energy region considered.

Concerning the I=1 amplitude, we showed that the situation is very similar to the one
of the $a_0(980)$ resonance, which is accepted as a resonance. Whether one decides to
call or not a resonance the $I=1$ pole that we find in an unusual
Riemann sheet, the resonant like
structure in the real axis has important repercussion in the photoproduction amplitudes and we can expect it to have important effects in many other observables. Since the relevant information is the I=1 amplitude in the real axis, the present work has provided values for this amplitude which can be tested in the study of future reactions.

 \section*{Acknowledgments}
  This work is partly supported by the Spanish Ministerio de Economia y Competitividad and European FEDER funds under the contract number
FIS2011-28853-C02-01, and the Generalitat Valenciana in the program Prometeo, 2009/090. We acknowledge the support of the European Community-Research Infrastructure Integrating Activity
Study of Strongly Interacting Matter (acronym HadronPhysics3, Grant Agreement
n. 283286) under the Seventh Framework Programme of EU.

\end{document}